\documentclass[%
 reprint,
superscriptaddress,
 amsfonts,amsmath,amssymb,amsthm,
 aps,
prl,
]{revtex4-1}

\usepackage{graphicx}
\usepackage{bm}
\usepackage{xcolor}
\usepackage{subfig}
\begin{document}



\title{Machine learning aided carrier recovery in continuous-variable quantum key distribution}

\author{Hou-Man~Chin}
\thanks{These two authors contributed equally}
\affiliation{Center for Macroscopic Quantum States (bigQ), Department of Physics, Technical University of Denmark, 2800 Kgs. Lyngby, Denmark}
\affiliation{Machine Learning in Photonic Systems group, Department of Photonics, Technical University of Denmark, 2800 Kgs. Lyngby, Denmark}
\author{Nitin~Jain}
\thanks{These two authors contributed equally}
\affiliation{Center for Macroscopic Quantum States (bigQ), Department of Physics, Technical University of Denmark, 2800 Kgs. Lyngby, Denmark}
\author{Darko~Zibar}
\affiliation{Machine Learning in Photonic Systems group, Department of Photonics, Technical University of Denmark, 2800 Kgs. Lyngby, Denmark}
\author{Ulrik~L.\ Andersen}
\email{ulrik.andersen@fysik.dtu.dk}
\affiliation{Center for Macroscopic Quantum States (bigQ), Department of Physics, Technical University of Denmark, 2800 Kgs. Lyngby, Denmark}
\author{Tobias~Gehring}
\email{tobias.gehring@fysik.dtu.dk}
\affiliation{Center for Macroscopic Quantum States (bigQ), Department of Physics, Technical University of Denmark, 2800 Kgs. Lyngby, Denmark}

\date{\today}

\begin{abstract}
The secret key rate of a continuous-variable quantum key distribution (CV-QKD) system is limited by excess noise. A key issue typical to all modern CV-QKD systems implemented with a reference or pilot signal and an independent local oscillator is controlling the excess noise generated from the frequency and phase noise accrued by the transmitter and receiver. Therefore accurate phase estimation and compensation, so-called carrier recovery, is a critical subsystem of CV-QKD. Here, we explore the implementation of a machine learning framework based on Bayesian inference, namely an unscented Kalman filter (UKF), for estimation of phase noise and compare it to a standard reference method. Experimental results obtained over a 20 km fibre-optic link indicate that the UKF can ensure very low excess noise even at low pilot powers. The measurements exhibited low variance and high stability in excess noise over a wide range of pilot signal to noise ratios. This may enable CV-QKD systems with low implementation complexity which can seamlessly work on diverse transmission lines.
\end{abstract}

\maketitle

Continuous-variable quantum key distribution (CV-QKD) enables information-theoretically secure key exchange between two parties using the continuous-variable properties of the quantized electromagnetic light field~\cite{Ralph2000, Grosshans2002, Diamanti2015, Laudenbach2018, Pirandola2019}. The quantum information used for generating the secret key can be imprinted onto coherent states in the amplitude and phase quadratures of laser light using electro-optical modulators at the transmitter. These quantum states are transmitted through an insecure channel -- typically assumed to be fully controlled by an adversary -- and measured by some form of coherent detection, e.g.\ radio-frequency heterodyne or phase-diverse homodyne detection at the reciever. The use of technology quite similar to that employed in classical coherent telecommunications~\cite{Kikuchi2015} is an attractive feature of CV-QKD with respect to integrability in existing telecom networks.
%

A CV-QKD coherent receiver uses a local oscillator (LO) to measure the quantum information carrying signal. 
Modern CV-QKD implementations generate the LO from a laser at the receiver, which is independent of the transmitter laser. This simplifies the CV-QKD implementation and increases security, however, at the cost of requiring to recover the frequency and phase of the quantum signal. This process, commonly known as carrier recovery in telecommunication \cite{Faruk2017}, is of utmost importance for the performance of CV-QKD implementations as an impairment cannot be distinguished from excess noise generated by an eavesdropper. 

The quantum signal operates in a significantly lower power regime than a typical optical telecommunications signal and correspondingly it is detected at a much lower signal-to-noise ratio (SNR). This regime is one in which traditional telecommunication algorithms for carrier recovery~\cite{Faruk2017} function quite poorly, if at all. Additionally, CV-QKD systems typically use a Gaussian modulation format that does not contain features present in traditional telecommunication formats, e.g. phase shift keying (PSK), which enable such algorithms to work. Therefore pilot-aided techniques in which a reference signal is transmitted together with the quantum signal have been developed and studied for CV-QKD systems~\cite{Qi2015, Soh2015, Huang2015, Marie2017}.

The `classical' reference and the quantum signal are usually time~\cite{Qi2015, Soh2015, Huang2015} or frequency multiplexed~\cite{Kleis2017,Brunner2019} and phase noise estimation is carried out on this reference, henceforth called the `pilot tone'. It is advantageous to have a pilot tone with as low power as possible to minimize interference with the quantum signal. Besides undesirable scattering effects in the fibre, a high power pilot tone has a negative effect on the signal to noise and distortion ratio of the digital-to-analog converter in the transmitter as well the analog-to-digital conversion in the receiver, thus decreasing the effective number of bits. To avoid some of these side effects the pilot tone and the quantum signal could be multiplexed in polarization, however, at the expense of (at least) doubling implementation complexity~\cite{Laudenbach2019}.
    
Here, we present a machine learning framework based on Bayesian inference, implementing an unscented Kalman filter (UKF) \cite{Sarkka2013} to estimate the phase of a pilot tone. The UKF's performance is investigated experimentally in a Gaussian-modulated CV-QKD protocol~\cite{Grosshans2002} operating over a 20\,km fibre link using an ultra-low linewidth laser and a standard telecommunications laser at the transmitter. The UKF achieves exceptional performance with excess noise figures below $1\%$ of the shot noise variance for a wide range of pilot tone SNRs. For instance, with the ultra-low linewidth laser, the UKF performs consistently well down to 3.5 dB pilot tone SNR ($\rm SNR_{pilot}$), which considerably relaxes the constraints on the filtering bandwidth. The UKF therefore not only enables higher secret key rates but also promises a more robust CV-QKD system with regards to environmental factors that may deteriorate $\rm SNR_{pilot}$. Moreover, it enables secret key generation using systems that would otherwise be unable to do so using the reference method.

Phase tracking in CV-QKD systems using Bayesian inference based on an extended Kalman filter or extended Kalman smoother has recently been studied on a CV-QKD system with a discrete modulation format using 8 coherent states \cite{Kleis2019}. In contrast to their implementation the UKF described here does not require careful fine tuning of algorithm parameters and is thus more stable under changing conditions. Gaussian modulation in comparison to discrete modulation has more mature security proofs~\cite{Leverrier2015}  but is more susceptible to phase noise (see Methods) because the optimum mean photon number of the transmitted quantum states is an order of magnitude higher. The UKF removes this as a significant limiting factor. Bayesian inference based methods have also been used for the measurement and characterization of laser phase noise, outperforming traditional methods in particular in the low laser power regime \cite{Brajato2019, Zibar2020}.

\section{Machine learning aided phase tracking algorithm for carrier recovery}
\begin{figure}[t]
\centering
    \subfloat[][]{
        \includegraphics[width=0.45\textwidth]{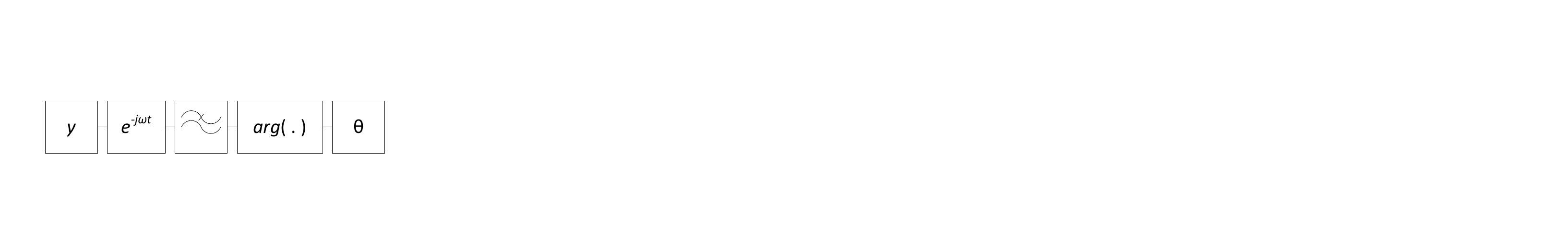}
        \label{fig:ref_dsp}}
       
    \subfloat[][]{
        \includegraphics[width=0.45\textwidth]{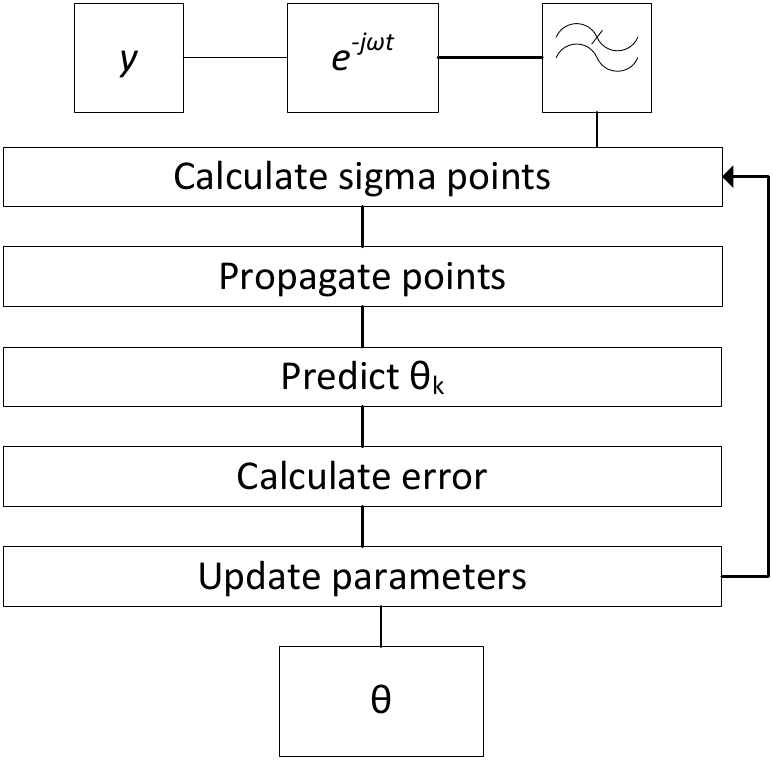}
        \label{fig:ukf_dsp}}
    \caption{Algorithms for (a) reference method and (b) machine learning approach}
    \label{fig:dsp}
\end{figure}

The phase noise associated with a time-varying pilot signal $y(t)$ acquired by a radio-frequency heterodyne receiver at discrete time instants $t = t_k$ can be corrected by evaluating 
\begin{equation}
    \theta_{k} \equiv \theta(t_k) = \tan^{-1}\bigg(\frac{\mathcal{H}(y(k))}{y(k)}\bigg)\ ,
\end{equation}
where $\mathcal{H}$ denotes the Hilbert transform. The linear trend in $\theta{_k}$ is removed to compensate for the frequency offset of the pilot tone leaving the phase noise. This method is standard for extracting the phase from a pilot signal and is equivalent to calculating the frequency offset, frequency shifting the pilot to baseband and then taking its argument, see Fig.~\ref{fig:ref_dsp}. In coherent detection systems the additive noise caused by the beating of the LO laser with vacuum fluctuations within the measurement bandwidth limit the efficacy of this method \cite{Zibar2019}. In principle, this can be solved by increasing the pilot signal power, however as previously mentioned, this may be undesirable in a practical CV-QKD system.

To overcome this pilot power limitation, we investigated a machine learning framework based on Bayesian inference. This approach allows inference of the phase from the noisy measurements $y_{k} := y(k)$. In theory such an approach is statistically optimal with respect to the mean square error when estimating the phase \cite{Sarkka2013}. To implement this method, a state space model that describes the evolution of the system in time is required in addition to a model that describes the measured values $y_{k}$. For the state space model, the phase of the quantum signal evolves with discrete Markovian dynamics and can be represented as

\begin{equation}
    X_{k} := \theta_{k} = \theta_{k-1} + q_{k-1}\ ,
    \label{eqn:ssmodel}
\end{equation}
where $X_{k}$ is the system state at symbol \textit{k}, $\theta_k$ is the phase at the same symbol and \textit{q} is the unknown (phase) process noise.
The measurement model of the pilot signal in a heterodyne detection system is given by a noisy measurement outcome $y_{k}$,

\begin{figure}[t]
\centering
	\includegraphics[width=0.45\textwidth]{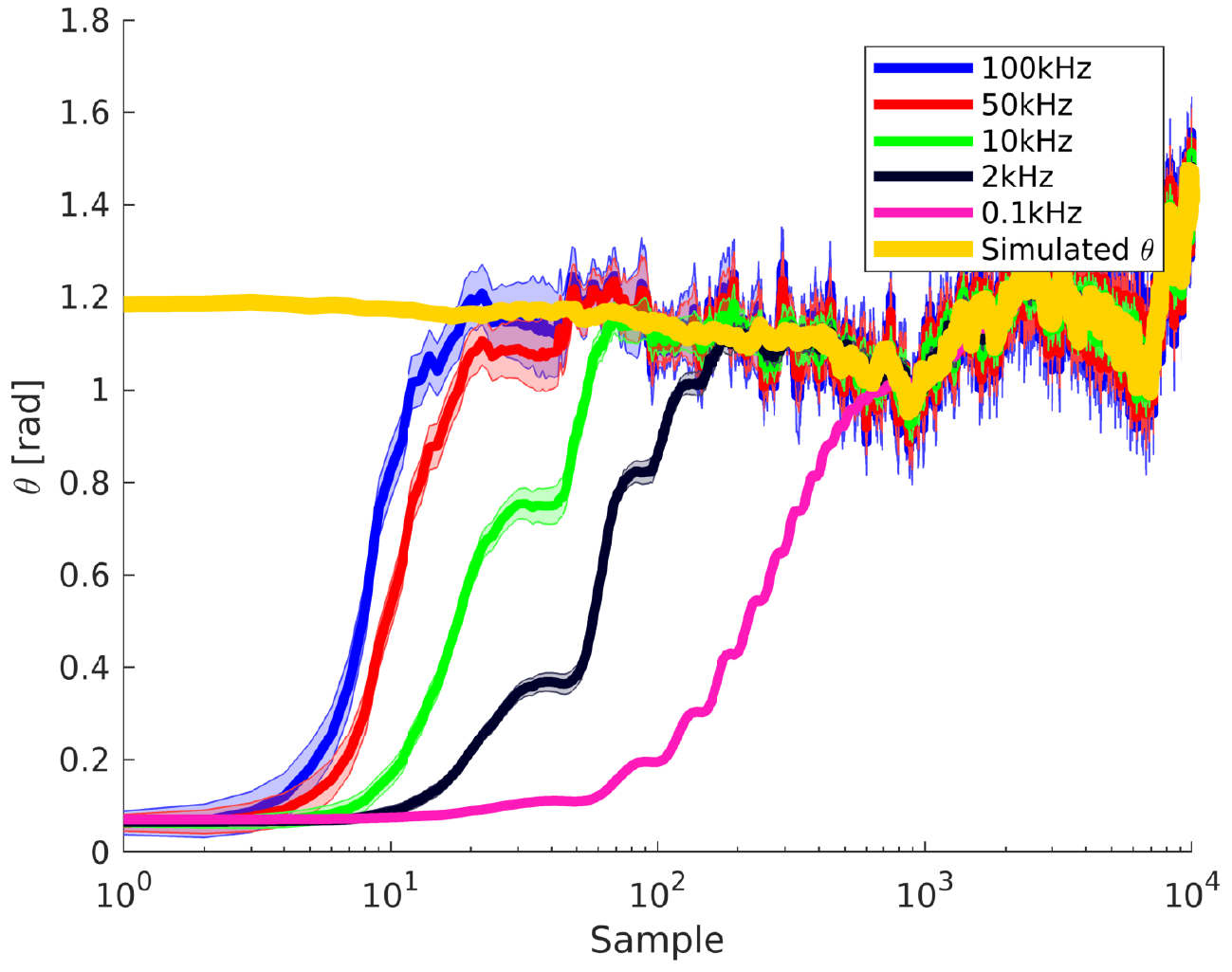}
	\caption{(Color online) UKF convergence performance with respect to (incorrect) laser linewidth input. The simulated phase noise stems from a 2 kHz linewidth laser. The tints around the traces indicate the standard deviation of the approximating Gaussian distributions used by the UKF.}
    \label{fig:ukf_convergence}
\end{figure}

\begin{figure*}[t]
    \hspace{-0.3cm}
    \subfloat[][]{
        \includegraphics[width=0.65\textwidth]{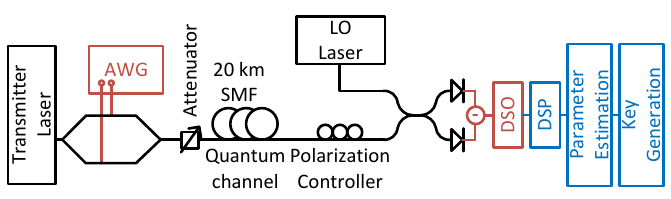}}%
    \hspace{-0.3cm}
    \subfloat[][]{
        \includegraphics[width=0.35\textwidth]{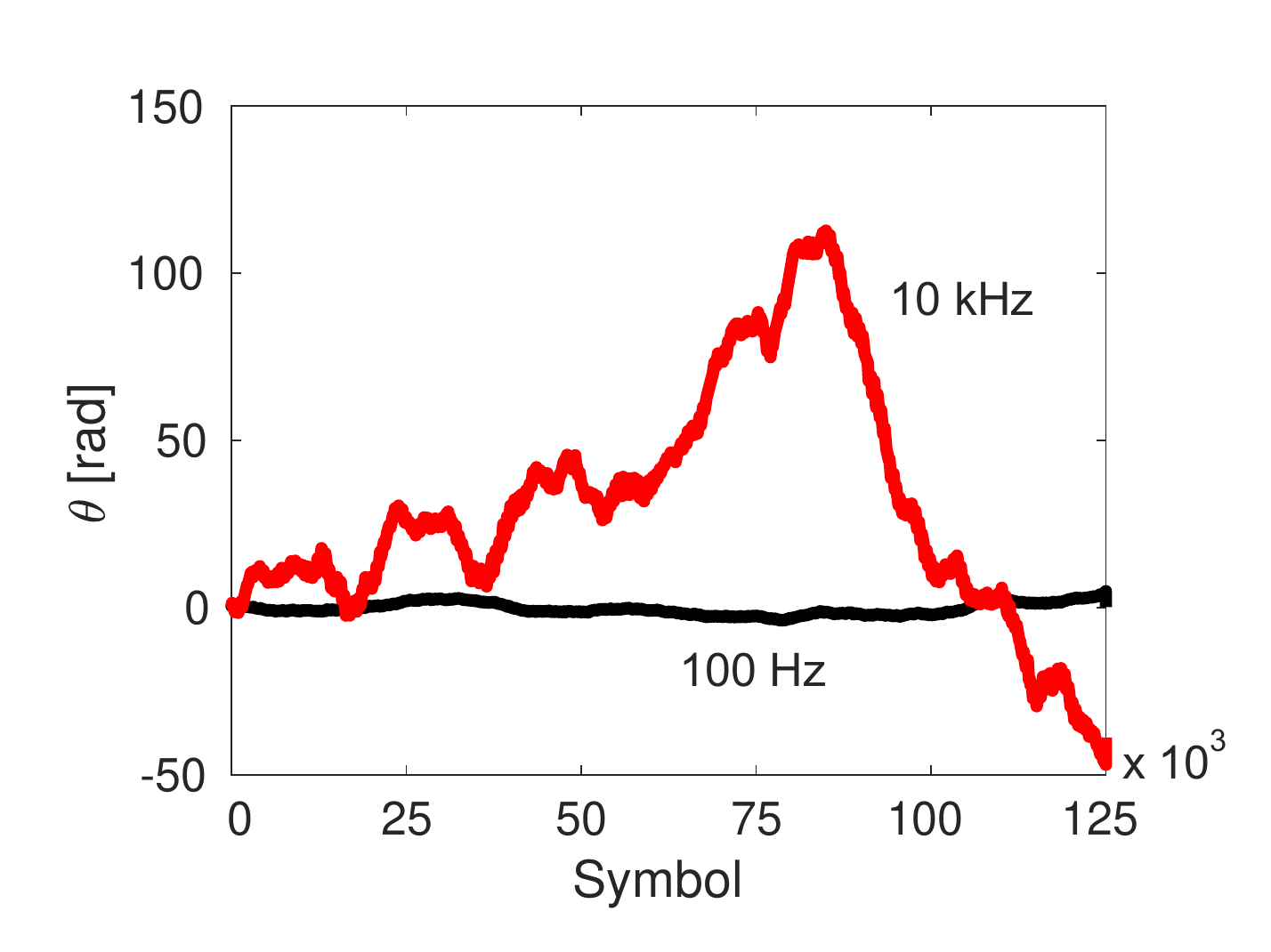}}%
    \caption{Experimental setup. a) An ensemble of coherent states at 1550\,nm was encoded into continuous-wave laser light by electro-optic in-phase/quadrature single sideband modulation driven by an arbitrary waveform generator (AWG). A reference pilot tone was digitally frequency multiplexed with these coherent states, which after suitable attenuation became the `quantum signal'. The polarization of the combined quantum signal and pilot tone, transmitted through a 20\,km SMF-28 fibre, was corrected with a manual polarization controller to match the polarization of an independent local oscillator (LO). The output of the radio-frequency heterodyne detector was sampled by a digital storage oscilloscope (DSO) at 1 GSamples/s before undergoing various digital-signal-processing (DSP) methods, including the unscented Kalman filter (UKF) assisted phase tracking. b) Sample phase profiles extracted by UKF at high $\rm SNR_{pilot}$ for 100 Hz and 10 kHz linewidth transmitter lasers. The receiver used the same $\approx$ 100 Hz linewidth laser as the LO for both setups.}
    \label{fig:setup}
\end{figure*}

\begin{equation}
    y_{k} = A\sin(\Delta \omega kT_{s} + \theta_{k}) + n_{k}\ ,
    \label{eqn:measmodel}
\end{equation}
where $A$ is the amplitude of the pilot signal, $\Delta \omega$ is the frequency offset between the LO laser and the pilot tone, $T_{s}$ is the sampling time granularity and $n_{k}$ is the shot noise corrupting the measurement. For each symbol $k$ Bayesian inference aims to obtain a filtering distribution
\begin{equation}
    p(\theta_{k}|y_{1:k})\ ,
\end{equation}
approximating $q$. The filtering distribution is the marginal distribution of the current $\theta_{k}$ given current and previous measurements $y_{1:k} = [y_{1},...y_{k}]$. The mean of this distribution is the statistically optimal estimated phase. 

A direct implementation of the problem can be intractable, and hence there are implementations of Bayesian inference which are less optimal but tractable. The UKF handles the non-linear system (Eqn.~\ref{eqn:measmodel}) by taking a Gaussian approximation of the process noise. As shown in Fig.~\ref{fig:dsp} (b), it does this by calculating some sigma points using the mean and standard deviation of the the approximating Gaussian distribution. These points are propagated through the measurement model which then are used to calculate the predicted mean and covariance. Similarly the mean and covariance of the measured noisy measurement are caluclated to estimate the error between the predicted state and the measured pilot. The Gaussian approximation is then updated using a Metropolis-Hastings algorithm and used to estimate the symbol phase. The updated distribution is then fed into the next symbol's estimation. In essence this allows for completely blind estimation without known system parameters. Should the given system parameters be significantly wrong, the major impact would be that convergence time to the optimum posterior distribution would be longer.

Figure~\ref{fig:ukf_convergence} shows the convergence time of the UKF when the initial process noise parameter (described by the laser linewidth) is varied for a simulated 2 kHz combined linewidth system. Underestimating the laser linewidth increases the convergence time of the UKF since underestimating limits the size of the steps the UKF can take towards the actual phase. This may restrict the UKF's ability to track the phase. Overestimating the linewidth can cause the UKF to overshoot as (barely) seen for the 100 kHz input but then settles to the system phase. The color tints on the figure show the standard deviation of the approximating Gaussian used by the UKF.

\section{Results}

\begin{figure*}
    \subfloat[][]{
        \includegraphics[width=0.48\textwidth]{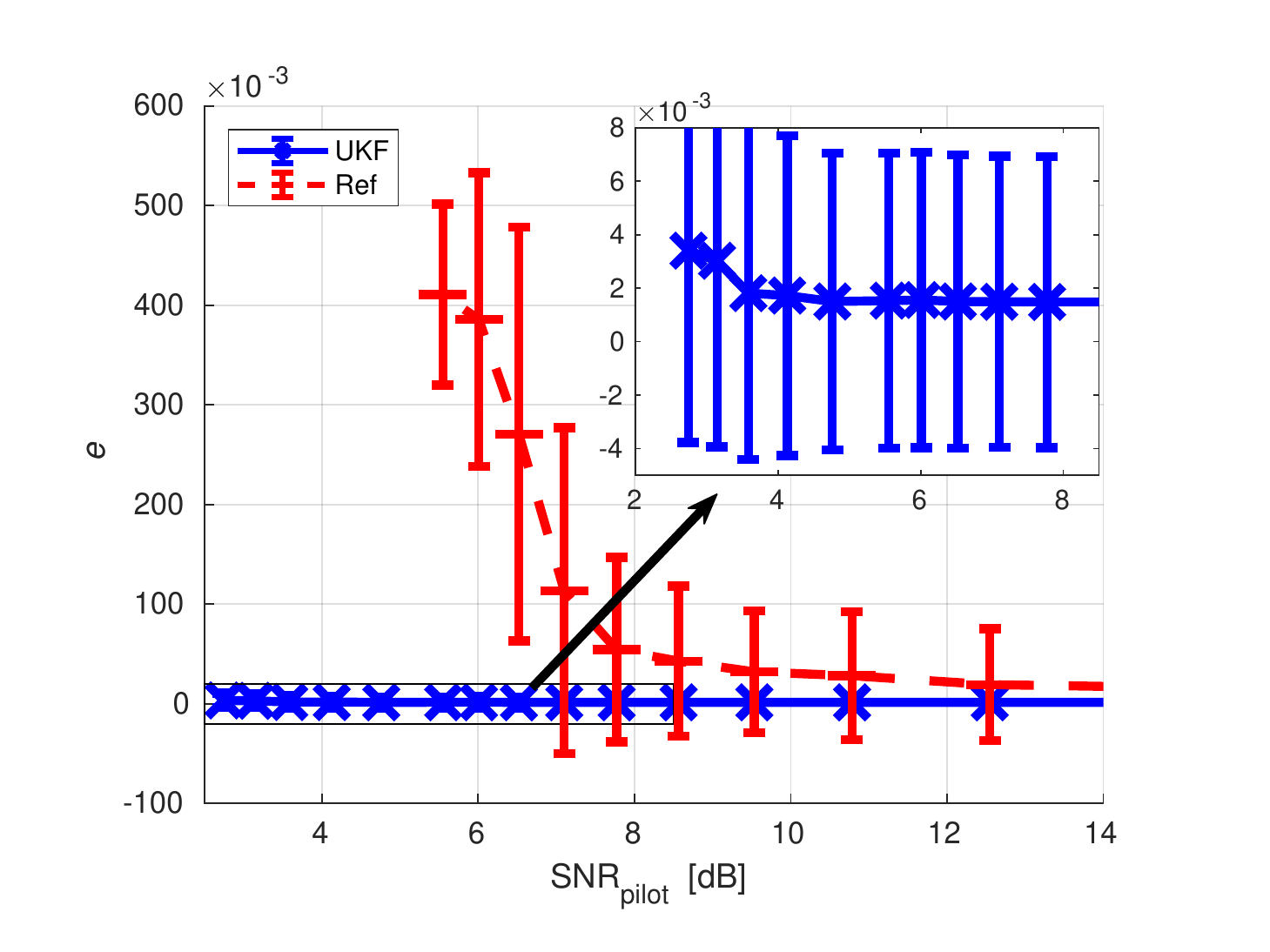}
        \label{fig:results1}}
    \subfloat[][]{
        \includegraphics[width=0.48\textwidth]{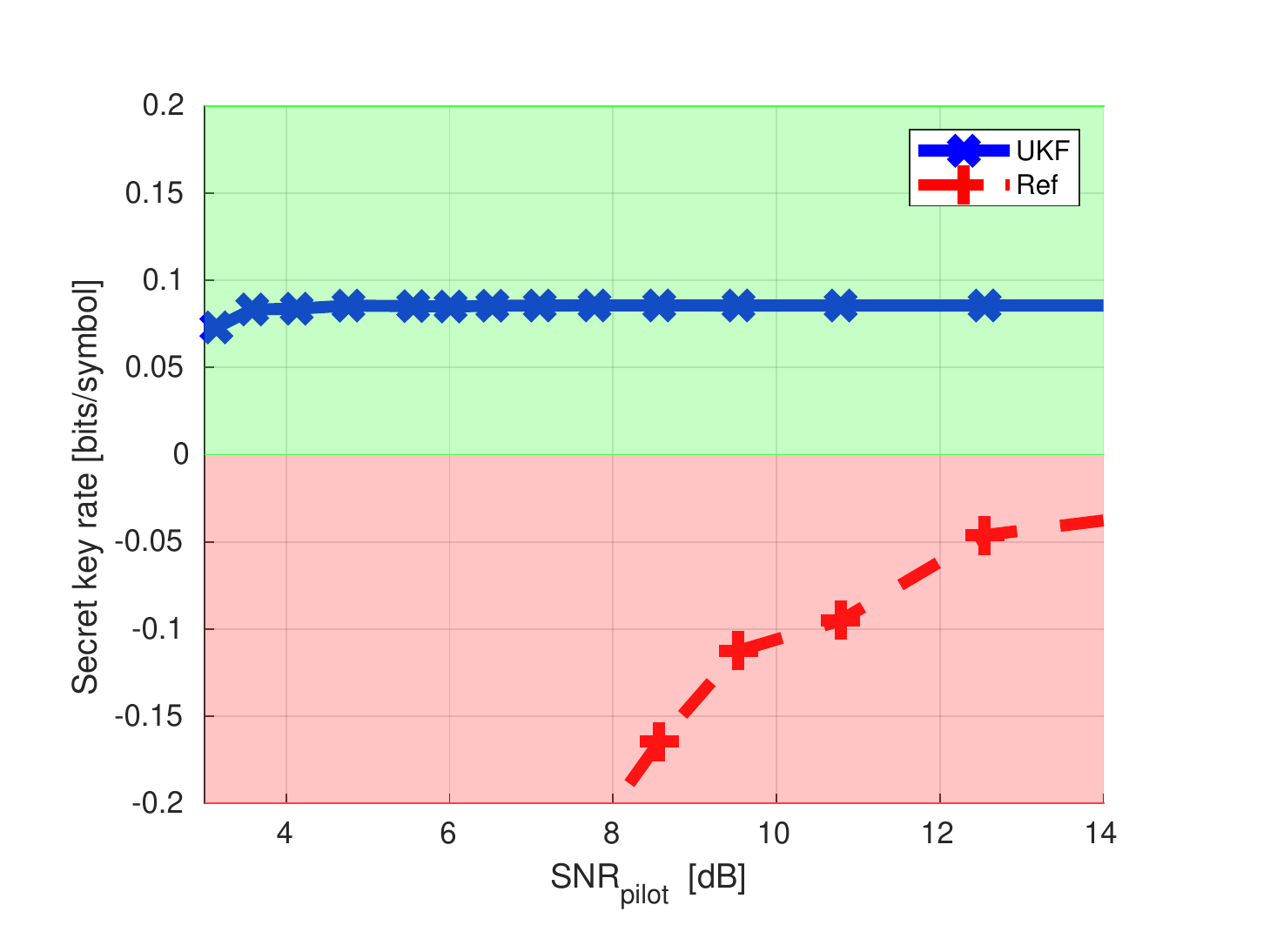}
        \label{fig:results2}}
        
    \subfloat[][]{
        \includegraphics[width=0.48\textwidth]{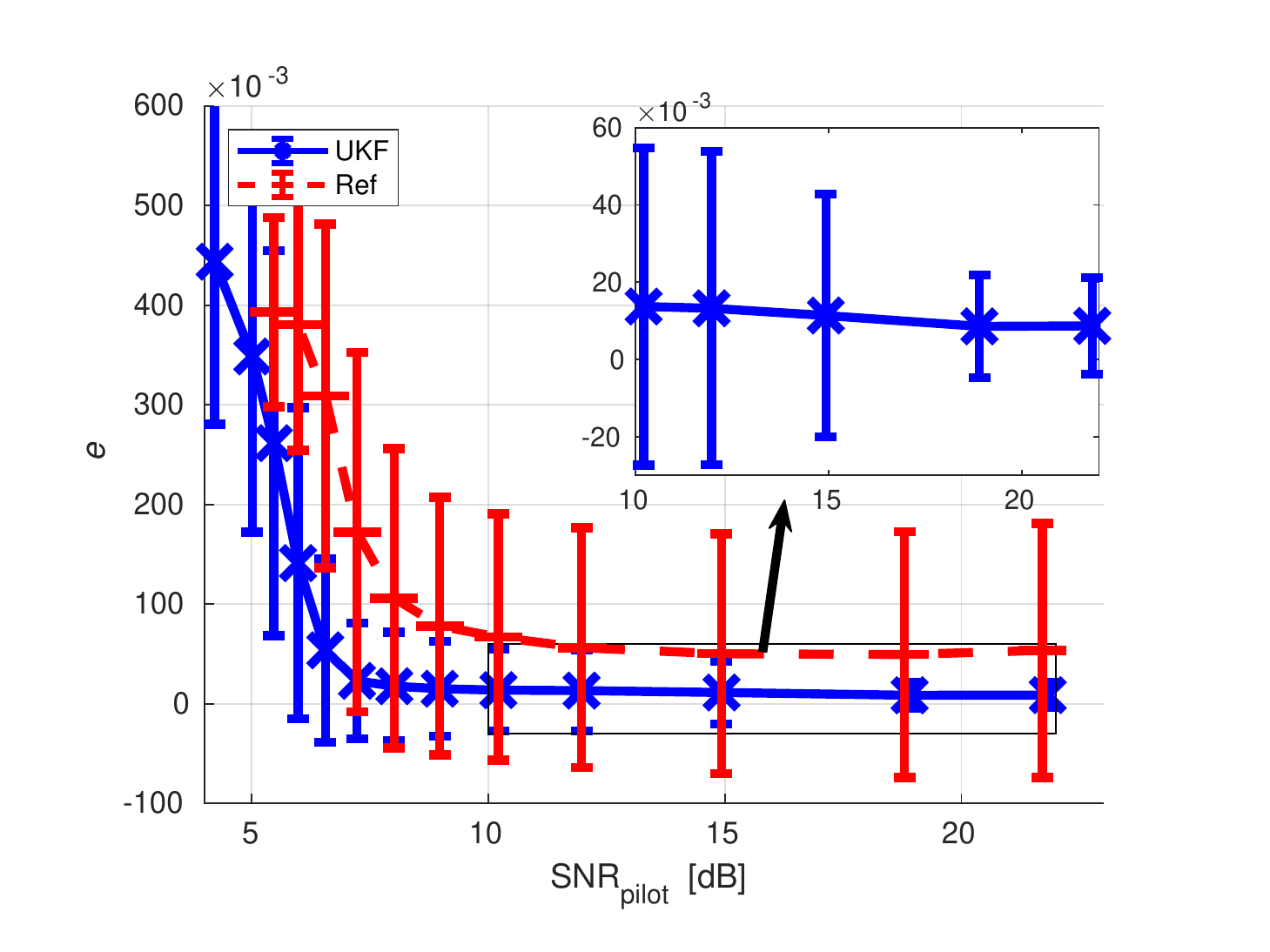}
        \label{fig:results3}}
    \subfloat[][]{
        \includegraphics[width=0.48\textwidth]{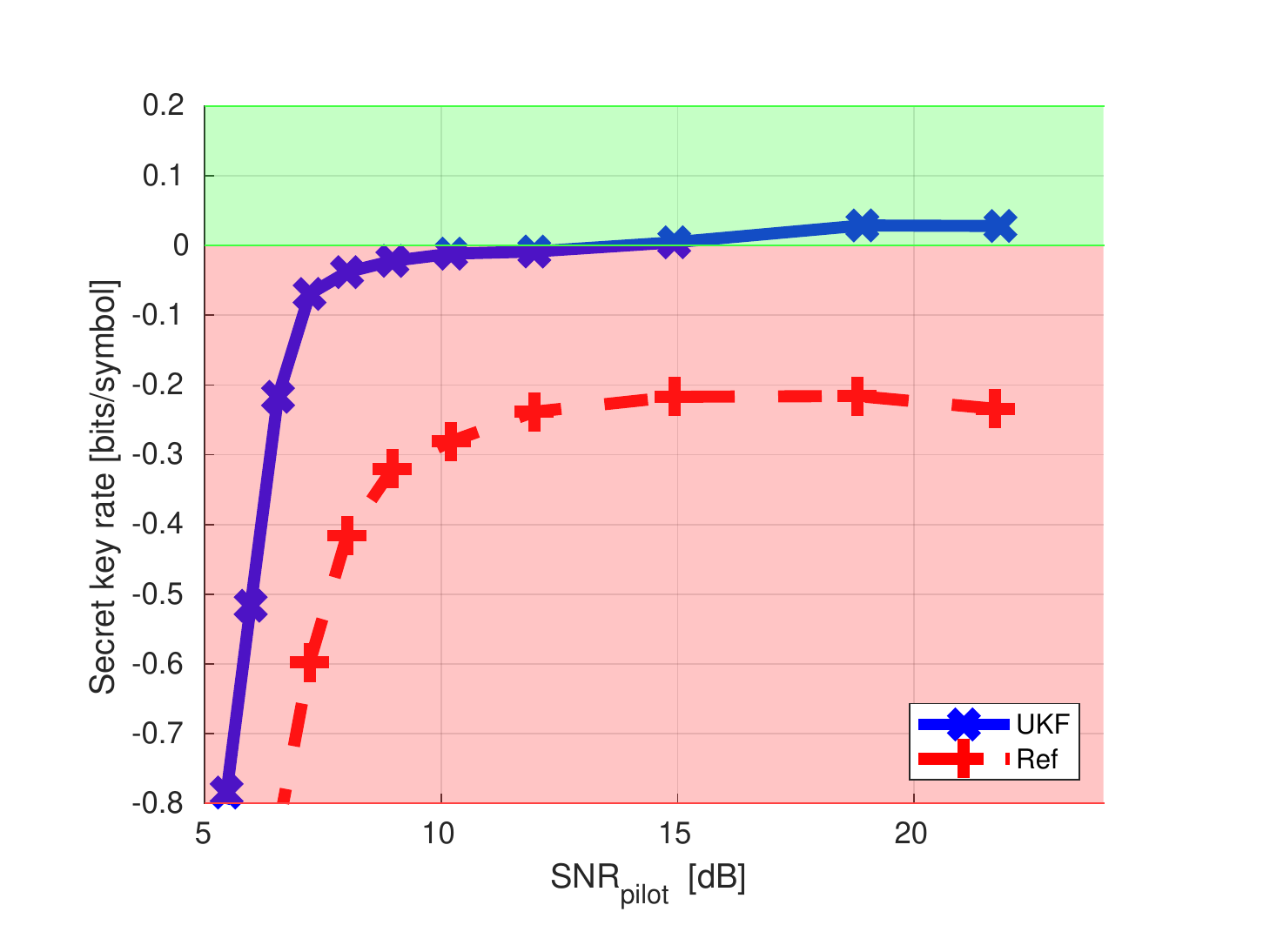}
        \label{fig:results4}}
    \caption{Experimental results demonstrating the UKF's performance. (a) Excess noise mean photon number $e$ obtained using both phase compensation methods and (b) respective estimated secret key rates. The thermal state at the transmitter's output had a mean photon number N = 2.73, the detector's electronic noise mean photon number $t\approx0.022$ was trusted and subtracted from $e$. We used the average value of $e$ and assumed an error reconciliation efficiency $\beta = 0.95$ in the key rate calculations. 
    The detector's optical efficiency of 0.84 was treated as trusted loss, i.e., not accessible to the eavesdropper. Error bars were calculated over 1000 frames. Experimental results using a 10 kHz laser in lieu of the 100 Hz laser in the transmitter. (c) Excess noise mean photon number for both phase compensation methods for N = 3.41 and (d) respective estimated secret key rates}
    \label{fig:results}
\end{figure*}

We investigated the algorithm's performance in a CV-QKD experiment, the setup for which is depicted in Fig.~\ref{fig:setup}a. The transmitter and receiver used commercially available telecom equipment and were connected by a 20\,km SMF-28 fibre channel. The transmitter prepared a 50 MBaud quantum signal and a frequency-multiplexed pilot tone, both inscribed into single sidebands of the electromagnetic field by an electro-optic in-phase and quadrature modulator. After suitable attenuation, the optical signal was sent to the receiver, either directly or through the 20\,km channel. Details about the modulation format at the transmitter and the coherent heterodyne detection at the receiver are described in the Methods. After acquiring the detection signal with an oscilloscope we performed several digital-signal-processing (DSP) steps (see Methods for further details), one of which is the proposed machine learning based carrier recovery algorithm, to recover the transmitted symbols. 

Using the transmitted and recovered symbols, we performed channel parameter estimation to obtain $e$, the excess noise mean photon number, and $\eta$, the combined optical efficiency of the transmission channel and the receiver's measurement device. We also estimated $N$, the mean photon number of the transmitted thermal state, with the transmitter and receiver connected directly. In Methods we describe in further detail how to estimate these parameters and calculate the achievable secret key rate. 

In the experiment we implemented two different transmitter lasers, a 100 Hz low linewidth fibre laser and a standard telecom external cavity diode laser with 10 kHz linewidth. The receiver's LO laser was always an identical model to the fibre laser. The lasers were free running, i.e.\ they were neither locked in frequency nor phase. Figure~\ref{fig:setup}b shows an example time trace of estimated phases. While for the fibre laser the phase varied only slightly over the course of 125k symbols (2.5 ms), the 10 kHz linewidth laser's phase drifted over a significantly larger range, requiring a larger standard deviation of the approximating Gaussian distribution. 

Figure~\ref{fig:results} a, c show the excess noise mean photon number $e$ versus $\rm SNR_{pilot}$ obtained from experimental measurements using the 100 Hz and the 10 kHz laser, respectively. Since the contribution to the excess noise caused by the electronic noise of the detector is assumed to be trusted, it was removed from the final result. The measurement set was divided into 1000 frames with 100k symbols per frame and channel parameter estimation was performed individually on each frame. The pilot tone SNR was varied by changing the filter bandwidth centred around the pilot tone frequency. This method was chosen in lieu of changing the pilot power at the transmitter to ensure a constant \textit{N} across different measurements and to isolate potential effects such as receiver saturation, optical non-linearity and bleeding of pilot power into the quantum signal that could have happened from adjusting the pilot power along a wide range. 

From the plots the superiority of the UKF is clear compared to the reference method. The inset in Fig.~\ref{fig:results}a shows that the UKF has no significant performance degradation using a 100 Hz linewidth laser at $\rm SNR_{pilot}$ as low as 4\ dB, with $e$ reaching $2 \times 10^{-3}$ at high $\rm SNR_{pilot}$. On the other hand, the reference method performs much worse than the UKF at low $\rm SNR_{pilot}$ and is still outperformed at the highest $\rm SNR_{pilot}$. Substituting in the 10 kHz linewidth laser gives overall worse results with the UKF deteriorating quickly at lower than 7 dB $\rm SNR_{pilot}$, though it still achieves $e < 0.01$ in the best case.

This is further put into perspective by the graphs in Fig.~\ref{fig:results}b,d that display the secret key rate calculated in the asymptotic regime using both phase compensation methods. Using the UKF it is always possible to extract a secret key using either transmitter laser, while even at a $\rm SNR_{pilot} = 26\,$dB, the reference method could achieve at best $e = 0.015$, which is still too high for key generation with the 100 Hz laser. For the 10 kHz linewidth laser, $e \approx 0.06$ was the best result. The worse performance may be due to the fast changing beat mode frequency of the lasers rendering the Gaussian approximation less accurate, however this requires further investigation.

Higher SNRs are limited by the pilot power in this experiment but theoretically the difference between the UKF and reference method should become negligible at sufficiently high SNR.
In fact, the reference method has been used for successful key generation~\cite{Kleis2017, Brunner2019, Laudenbach2019} albeit this was for discrete modulations formats and/or different experimental settings.

\section{Conclusion}
This work shows the performance increase achieved by employing a machine learning Bayesian inference framework for the compensation of laser phase noise in a Gaussian modulation CV-QKD setup operating over a distance of 20 km. Using a relatively low pilot power the machine learning approach enabled secret key generation in our system for two very low linewidth lasers (100 Hz) as well as for a system using one comparatively larger linewidth laser (10 kHz).
The demonstrated performance is consistent over $\text{SNR}_\text{pilot}$ range exceeding 10 dB. Future CV-QKD systems operating in telecom networks that use fibres with varying attenuation and noise, e.g.\ stemming from wavelength division multiplexed data transmission, would experience a degradation of the available $\text{SNR}_\text{pilot}$. In such environments, the UKF may be the only method that guarantees key generation without having to adapt the pilot power. Finally, given the moderate symbol rates employed in CV-QKD, real-time implementations of the UKF should be feasible, thus making it a substantial element in all CV-QKD systems that implement the LO using an independent laser.

\section*{Methods}
\subsection{Experimental setup}
\label{sec:setup}
The experimental setup used to perform CV-QKD is shown in Fig.~\ref{fig:setup}. The transmitted symbols were drawn from independently seeded pseudo random Gaussian distributions with variance of 1 and zero mean at a rate of 50 MBaud. These digital symbols were upsampled to the 500 MSamples/s sampling rate of the arbitrary waveform generator (AWG) after which they were frequency shifted to $\Omega_q = 60$ MHz, i.e. multiplied with $\exp(j\Omega_q t)$, for single sideband modulation. A reference pilot tone at a frequency of $\Omega_p = 130$ MHz was also multiplexed with the quantum signal for the purpose of phase noise compensation and frequency offset estimation. This radio frequency signal and a $\pi/2$-phase shifted version thereof drove the two arms of an I-Q electro-optical modulator to simultaneously modulate the quantum signal in both quadratures onto the output of laser centered at  $1550.13\,$nm. The optical signal was then attenuated such that the mean photon number from only the quantum signal (i.e. excluding the pilot tone) was $\approx$ 2.73 at the quantum channel input.

At the channel output, the transmitted optical signal was detected using a balanced heterodyne coherent receiver with a free-running LO generated by laser separate from the transmitter with an offset frequency $\approx$ 200\, MHz. 
The LO power was 9\,dBm, giving a shot noise clearance of $\approx$ 13\,dB over the electronic noise. The output of the balanced receiver passed through a 200\,MHz low pass filter and was then digitized by a 10 bit digital storage oscilloscope (DSO) whose clock synchronized to that of the AWG to avoid additional penalties from clock recovery algorithm. The optical efficiency of the balanced detector (due to the non-unity quantum efficiency of the photodiodes and optical loss from connectors) was measured to be $\approx 0.84$.

The measurement time was divided into frames, each consisting of 100k complex values, or the `quantum symbols'. A 10k symbol long CAZAC sequence~\cite{Heimiller1961}, appended to the quantum symbol sequence, aided in timing recovery, synchronization and bulk phase offset compensation. 
\subsection{Digital-signal-processing (DSP)}
\label{sec:DSP}
Additional DSP is performed to facilitate QKD system operation. The transmitted quantum symbols are shaped with a root raised cosine filter with roll-off of 0.4 and matched filtering is performed at the receiver. The pilot signal is filtered using a wide band pass filter centred on its approximate location, calculated through the power spectrum of the received signal. The frequency offset is estimated through a Hilbert transform of the pilot and a linear fit of the extracted phase profile. This is re-estimated once more using the desired bandwidth filter which then shifts the pilot to baseband using the frequency offset estimate. The time varying phase is left when taking the argument of the pilot tone. Note that this baseband pilot is the input to the UKF after downsampling to symbol rate. The received quantum signal, vacuum and electronic noise calibration measurements were whitened with respect to the measurement of the combined vacuum and electronic noise power spectrum.

\subsection{Excess noise and secret key rate calculations}
\label{sec:skr}
\begin{figure}
    \centering
    \includegraphics[width=0.4\textwidth]{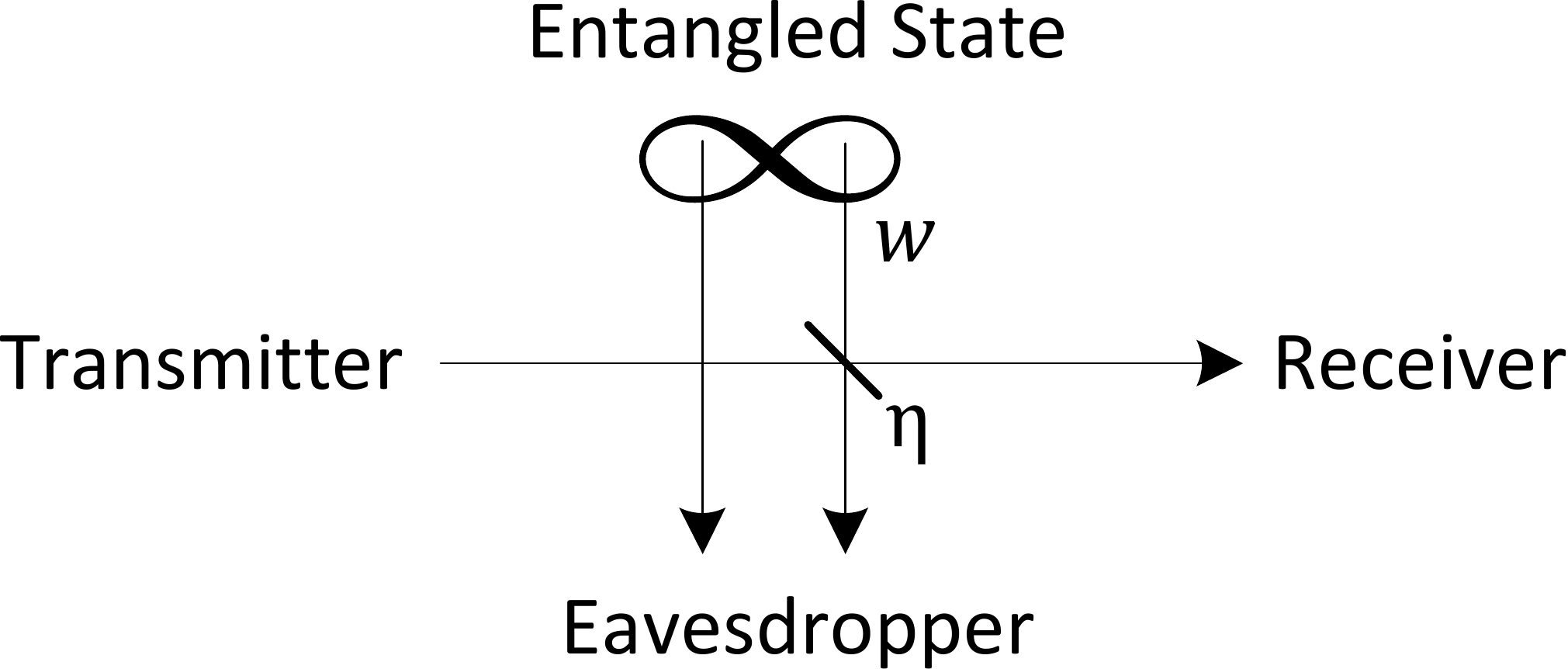}
    \caption{Channel model based on the entangling cloner attack. An eavesdropper injects one mode of an entangled state with mean photon number $w$ into the open port of a beam splitter with transmittance $\eta$ describing the optical channel loss. The excess noise mean photon number at the channel output is $e = w(1-\eta)$. }
    \label{fig:ChannelModel}
\end{figure}
To quantify the performance we use the secret key rate achievable in the asymptotic limit as well as the excess noise mean photon number at the channel output following an entangling-cloner attack model as depicted in Fig.~\ref{fig:ChannelModel}. The prepare-and-measure covariance matrix between the symbols chosen from a Gaussian probability distribution at the transmitter and measurement outcomes from a heterodyne (or phase diverse) receiver is
\begin{align}
\gamma =     
\left(\begin{matrix}
    2N & 0 & N\sqrt{2 \eta} & 0\\
    0 & 2N & 0 & N\sqrt{2 \eta }\\
    N\sqrt{2 \eta} & 0 & N \eta + e + 1 & 0\\
    0 & N\sqrt{2 \eta} & 0 & N \eta + e + 1
    \end{matrix}\right)\ ,
\end{align} 
where $N$ is the mean photon number of the transmitted thermal state, $e$ is the excess noise mean photon number at the transmission channel output, $\eta$ is the combined optical efficiency of the transmission channel and the receiver's measurement device~\cite{Laudenbach2019}. In a practical CV-QKD implementation the covariance matrix is estimated from the symbols as follows.
\begin{align}
\hat{\gamma} =     
\left(\begin{matrix}
    2N & 0 & \hat{z} & 0\\
    0 & 2N & 0 & \hat{z}\\
    \hat{z} & 0 & \hat{y} & 0\\
    0 & \hat{z} & 0 & \hat{y}
    \end{matrix}\right)\ .
\end{align} 
It is assumed that the transmitted thermal state has been previously characterized, i.e.\ $N$ is known. The parameters $\eta$ and $e$, inferred from the estimated covariance matrix as
\begin{align}
    \hat{\eta} &= \frac{\hat{z}^2}{2N^2}\ ,\\
    \hat{e} &= \hat{y} - \frac{\hat{z}^2}{2N} - 1\ ,
\end{align}
give the asymptotic secret key rate, 
\begin{equation}
    K = \beta I(A:B) - S(B:E)\ .
\end{equation}
Here, $A, B, E$ denote the modes of the transmitter, receiver, and eavesdropper, respectively, $I$ is the mutual information, $S$ is the Holevo information and $\beta$ is the information reconciliation efficiency. For the sake of simplicity, we ignore finite-size effects here but more details can be found in \cite{Leverrier2015}. 

Phase noise stemming from imperfect phase tracking effectively reduces the covariance term $\hat{z}$ by a factor $\kappa = \exp(-\sigma_\text{pn}^2 / 2)$, assuming Gaussian-distributed phase noise with $\sigma_\text{pn}$ as the standard deviation. If the phase noise is untrusted, i.e.\ $\kappa$ is unknown and unaccounted for, we obtain (via the entangling cloner model) a reduction of the actual physical transmittance of the channel to a virtual one, $\eta^\prime = \kappa^2\eta$. Simultaneously, the increased excess noise is given by $e^\prime = e + (1-\kappa^2)N\eta$. Thus, the larger the mean photon number of the ensemble of coherent states, the larger the effect of the phase noise.

\section*{Acknowledgment}

The authors gratefully acknowledge support by the European Research Council through the ERC-CoG FRECOM project, the Danish National Research Foundation, Center for Macroscopic Quantum States (bigQ, DNRF142) and SPOC Research Center of Excellence and EU project CiViQ (grant agreement no.\ 820466). 

\bibliographystyle{unsrt2authabbrvpp}

\bibliography{bibliography}

\end{document}